\input{aipcheck}

\documentclass[
    ,final            % use final for the camera ready runs
  ]
  {aipproc}

\layoutstyle{6x9}

\begin{document}

\title{The Observational Signatures of Primordial Pair-Instability Supernovae}


\author{Daniel J. Whalen}{
  address={Department of Physics, Carnegie Mellon University, Pittsburgh, PA  15213 }
}

\author{Chris Fryer}{
  address={Los Alamos National Laboratory, Los Alamos, NM  87545}
}

\begin{abstract}

Massive Population III stars from 140 - 260 M$_{\odot}$ ended their lives as 
pair-instability supernovae (PISNe), the most energetic thermonuclear explosions 
in the universe.  Detection of these explosions could directly constrain the 
primordial IMF for the first time, which is key to the formation of the first 
galaxies, early cosmological reionization, and the chemical enrichment of the 
primeval IGM.  We present radiation hydrodynamical calculations of Pop III  
PISN light curves and spectra performed with the RAGE code.  We find that the 
initial radiation pulse due to shock breakout from the surface of the star, 
although attenuated by the Lyman-alpha forest, will still be visible by JWST 
at $z \sim$ 10 - 15, and possibly out to $z \sim$ 20 with strong gravitational 
lensing.  We have also studied metal mixing at early stages of the explosion 
prior to breakout from the surface of the star with the CASTRO AMR code and
find vigorous mixing in primordial core-collapse explosions but very little in 
PISNe. This implies that the key to determining progenitor masses of the first 
cosmic explosions is early spectroscopy just after shock breakout, and that 
multidimensional mixing is crucial to accurate low-mass Pop III SNe light 
curves and spectra. 

\end{abstract}

\maketitle

%%%%%%%%%%%%%%%%%%%%%%%%%%%%%%%%%%%%%%%%%%%%
%% MAINMATTER
%%%%%%%%%%%%%%%%%%%%%%%%%%%%%%%%%%%%%%%%%%%%

\section{Introduction}

The masses of the first stars, or the primordial IMF, is key to the character 
of the first galaxies, the chemical enrichment and reionization of the early 
IGM, and the origin of the supermassive black holes found at the centers of 
most massive galaxies today.  No direct observation of Pop III stars can yet 
constrain the Pop III IMF.  However, detection of the first cosmic explosions 
could determine the masses of their progenitors, and thereby the Pop III IMF.  
Stellar evolution models predict that 15 - 50 M$_{\odot}$ stars die in core
collapse supernovae (SNe) and that 140 - 260 M$_{\odot}$ stars die in much 
more energetic PISNe \citep{hw02}. We have performed radiation hydrodynamical 
simulations of the light curves and spectra of primordial PISNe in anticipation 
of their discovery by JWST and the next generation of transient detection 
satellites.

\section{Simulations}

We performed a suite of ten numerical simulations with the RAGE code, an 
Eulerian adaptive mesh refinement radiation hydrodynamics code developed 
at Los Alamos National Laboratory (LANL) \citep{rage08}.  We considered 
five zero-metallicity and five $Z =$ 10$^{-4} Z_{\odot}$ progenitors with 
masses of 150, 175, 200, 225 and 250 M$_{\odot}$.  The progenitor stars 
were first evolved in the 1D implicit hydrodynamics code KEPLER and then 
exploded. The blast was evolved for 100 s in KEPLER, until the end of all 
nuclear burning. These blast profiles were then mapped onto a 1D spherical 
coordinate AMR grid in RAGE and evolved out to three years, the typical 
lifetime of PISN light curves (LCs).  The LCs of PISNe persist this long 
because radiation diffusion timescales through the their massive ejecta are 
on the order of a year.  In our models we used grey flux-limited diffusion 
(FLD) radiation transport, LANL TOPS atomic opacities, and two-temperature 
(2T) physics in which radiation and matter are coupled but not assumed to 
be at the same temperature.

\begin{figure}
  \includegraphics[height=.3\textheight]{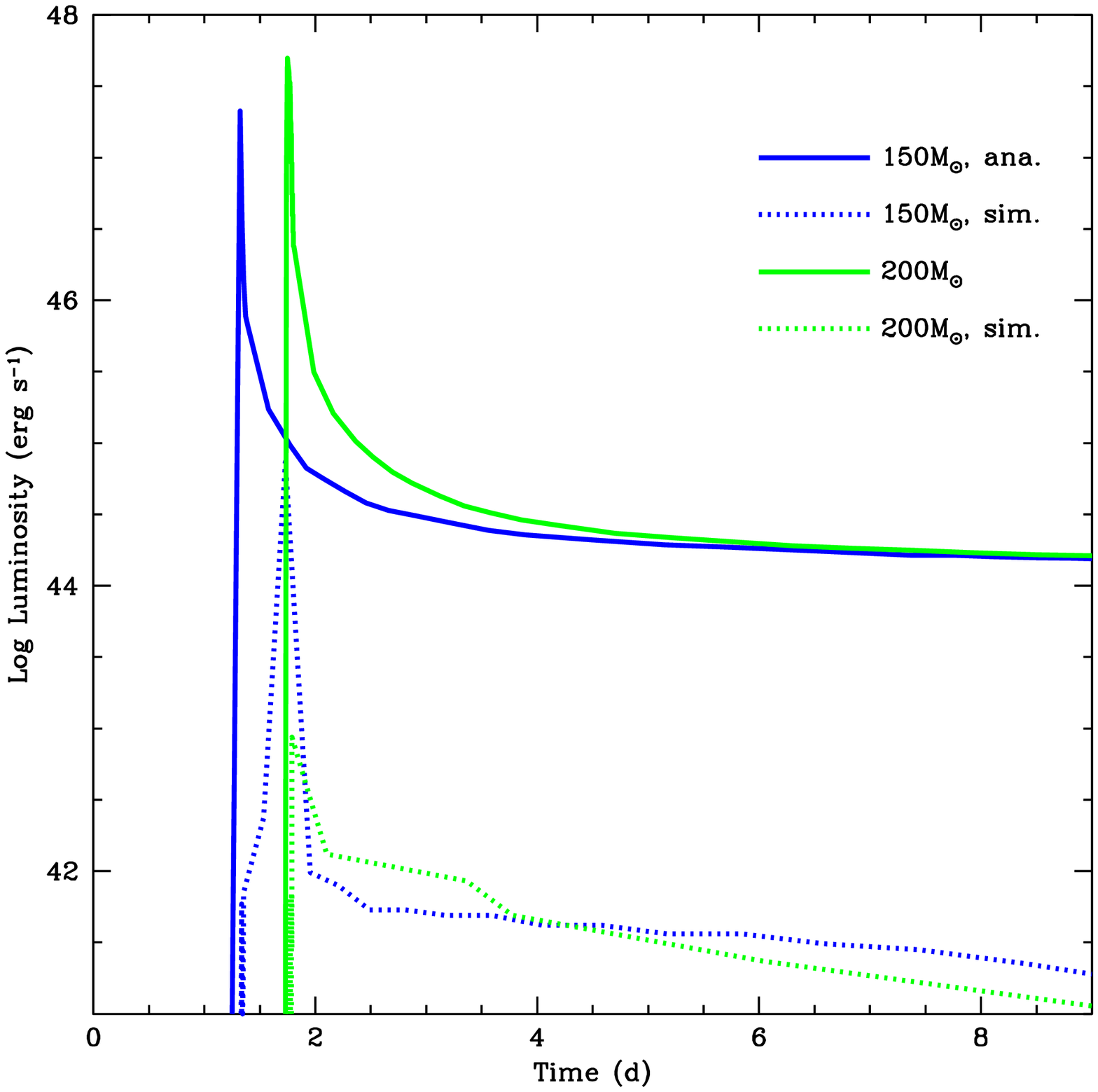}
  \includegraphics[height=.3\textheight]{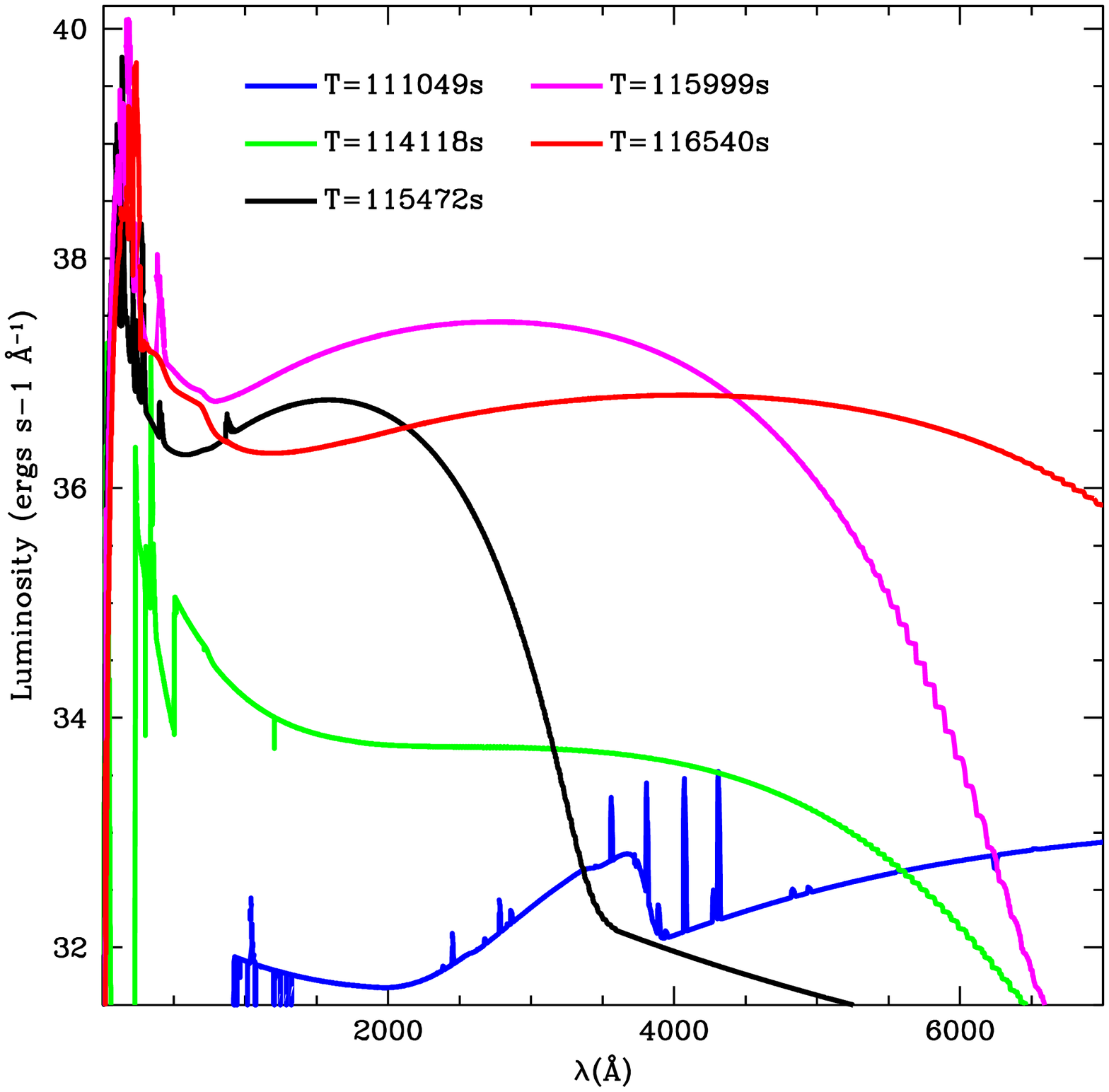}
  \caption{{\it Left:} bolometric light curves (luminosity vs. time) of 
shock breakout (dark line: 150\,M$_\odot$ model, light line: 200\,M$_\odot$) 
as predicted by semianalytical models (solid) and by our full 2T radiation 
hydrodynamics calculation (dotted).  Note that the peak in calculated 
emission is over an order of magnitude less than that of the simple estimate.  
{\it Right:} spectra (luminosity vs. wavelength) for a 150\,M$_\odot$ PISN at 
5 times during shock breakout.  
}

\end{figure}

\section{PISN Light Curves and Spectra}

In the left panel of Figure 1 we show bolometric light curves of shock 
breakout for 150 and 200 M$_{\odot}$ PISNe predicted by both semianalytical
models and our 2T RAGE calculations. The semianalytical curve is constructed
by first computing the position of the photosphere in successive snapshots 
of the RAGE hydrodynamical profiles assuming an elecron-dominated opacity
ahead of the shock, similar to what has been done in past studies 
\citep{scan05}.  The bolometric luminosity was then calculated taking the
spherical shock to be a blackbody source at this temperature and position. 
In our simulated luminosity, we compute the true emission with radiation
transport and frequency-dependent TOPS opacities.  Note that the peak in 
calculated emission is over an order of magnitude less than that of the 
simple estimate, indicating that accurate reproduction of shock breakout 
mandates full radiation hydrodynamics.  In the left panel of Figure 1 we 
show spectra (luminosity vs. wavelength) for a 150\,M$_\odot$ PISN at 5 
times during shock breakout.  A broad set of emission and absorption lines 
is present during this process. The initial burst of photons is at shorter 
wavelengths, 0.1 - 1 keV. As expected, as the shock expands and cools the 
emission at low wavelengths increases.       

\begin{figure}
  \includegraphics[height=.35\textheight]{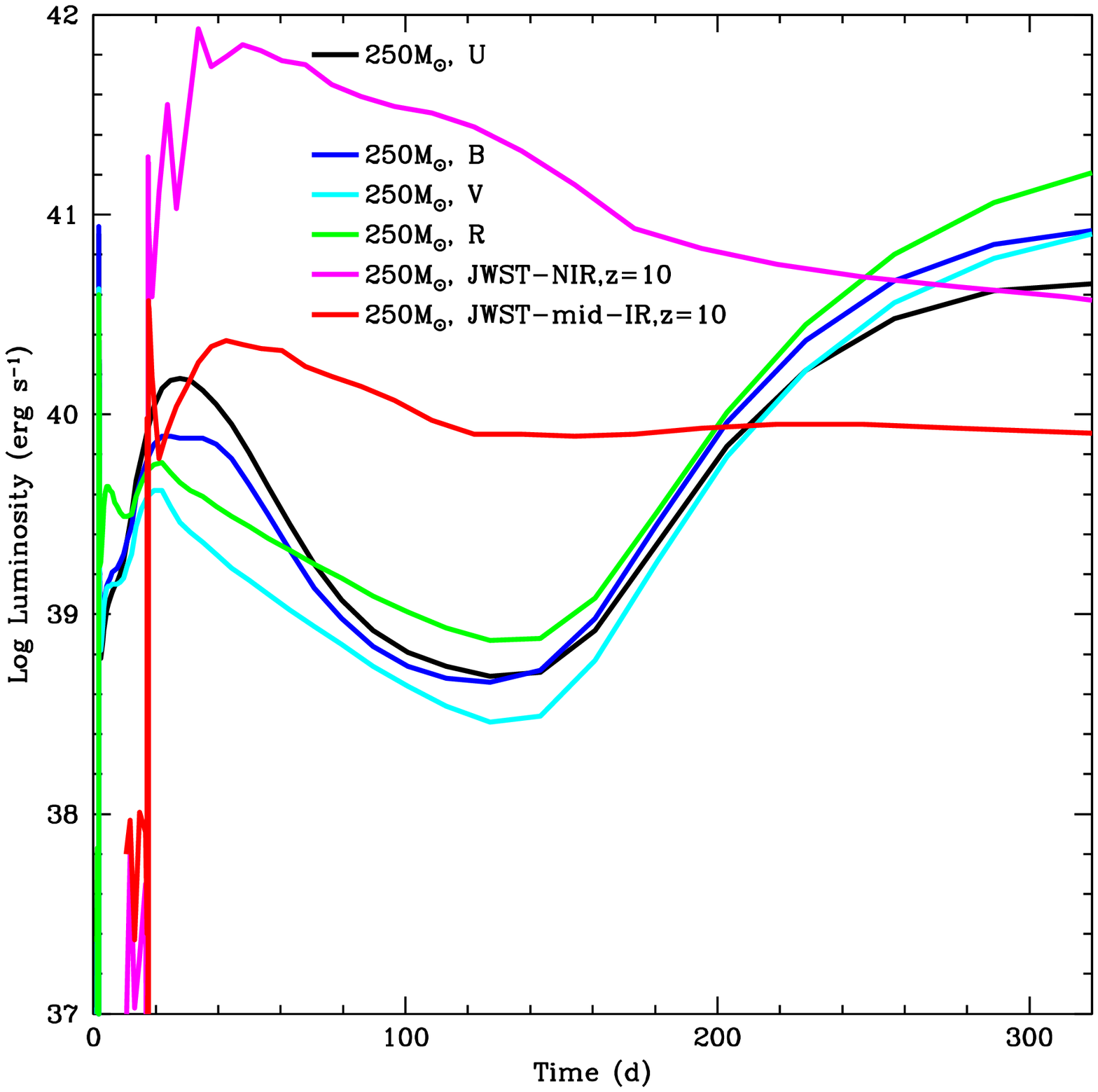}
  \includegraphics[height=.35\textheight]{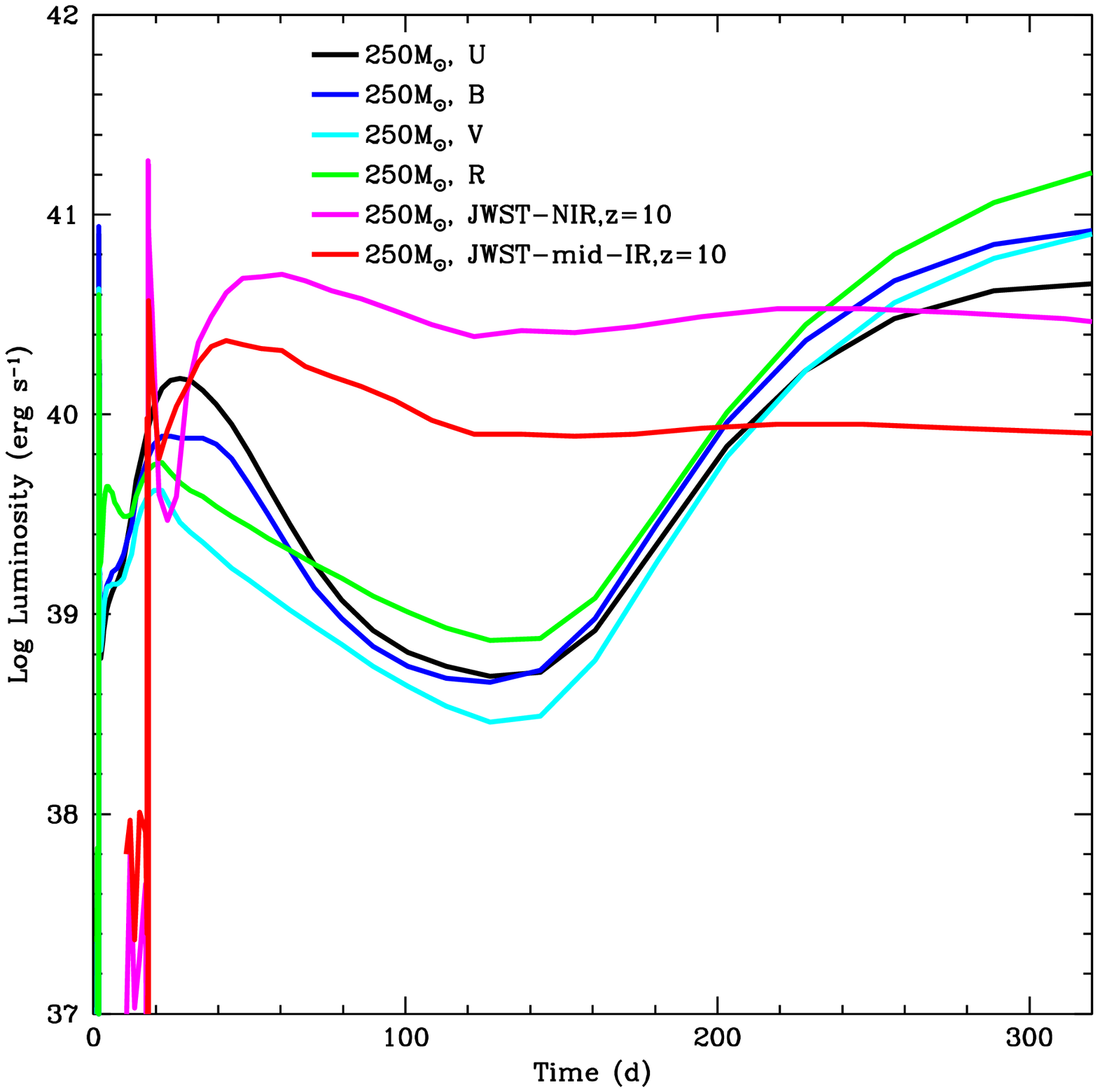}
  \caption{Light curves (luminosity versus time) for several spectral bands 
for a 250 M$_{\odot}$ PISN out to 1 year.  The differences between the plots 
on the left and on the right are due to the inclusion of the Lyman break, 
which cuts off any emission shortward of 1216 \AA.  Most high-redshift SN 
searches for these PISNe will detect the shock breakout.}
\end{figure}

We show light curves for several spectral bands for a 250 M$_{\odot}$ PISN 
out to 1 yr in Figure 2. In the left panel we include two bands consistent 
with the JWST bandpass, assuming that the source is at $z \sim$ 10. Here, we 
account both the wavelength shift of the source emission and time dilation.  
Note that the shock breakout emission lasts for a week in the JWST frame. 
Several of the bands exhibit the classical initial peak and steady decline 
over 300 days in the rest frame but the NIR and mid-IR flux dips and then 
rises at late times.  

What most distinguishes Pop III PISN detection from current Type Ia searches 
is the Lyman alpha forest.  Most of the flux shortward of 1216 \AA\ will be 
scattered out of the line of sight of the explosion by neutral hydrogen that 
has not yet been ionized by high-$z$ star-forming galaxies and quasars.  We 
estimate the component of the source signal that would actually reach us by 
excluding all photons shortward of the Lyman limit, which we show in the 
right panel of Figure 2. Although the breakout flux is attenuated by a factor 
of ten by Lyman-$\alpha$ scattering, the residue still peaks at above 10$^{41}
$ erg s$^{-1}$, which our initial estimates indicate will be visible to JWST 
out to $z \sim$ 10 - 15.  Strong lensing may extend this detection limit to $z 
\sim$ 20 (Holz, Whalen \& Fryer in prep).

Because PISN are visible for 30 - 60 years in the rest frame, they will most
likely be detected by their initial breakout pulse.  It is not yet clear if
future cosmic transient finders such as the successor to Swift will have the
sensitivity to detect such explosions.  However, they will be detected by
JWST during deep-field surveys of $z >$ 10 protogalaxies, with spectroscopic 
followup from large 30-meter class telescopes on the ground.  Because CC SNe
and PISNe have similar breakout LCs, early spectroscopy is the best bet for 
discriminating between them because heavy elements may appear sooner in CC SNe 
spectra.  In Figure 3 we show 2D CASTRO AMR simulations of the distribution of 
elements in both CC SNe (left) and PISNe (right) in the star before the shock 
has reached the surface \citep{joggerst10}. There is vigorous mixing of heavy 
elements from deep in the interior of the star to its outer layers in the CC 
case but not in the PISN. If these heavy elements appear in the emission 
spectra of the light curve just after shock breakout they would be a clear 
signature of a low-mass Pop III progenitor rather than a very massive one.  
Finally, we note that PISN remnants enclose large volumes of the IGM at
later times, eventually depositing up to 50\% of the original energy of the 
explosion into CMB photons that may be detectable as excess power at small 
scales in the CMB \citep{wet08b}.

\begin{figure}
  \includegraphics[height=.30\textheight]{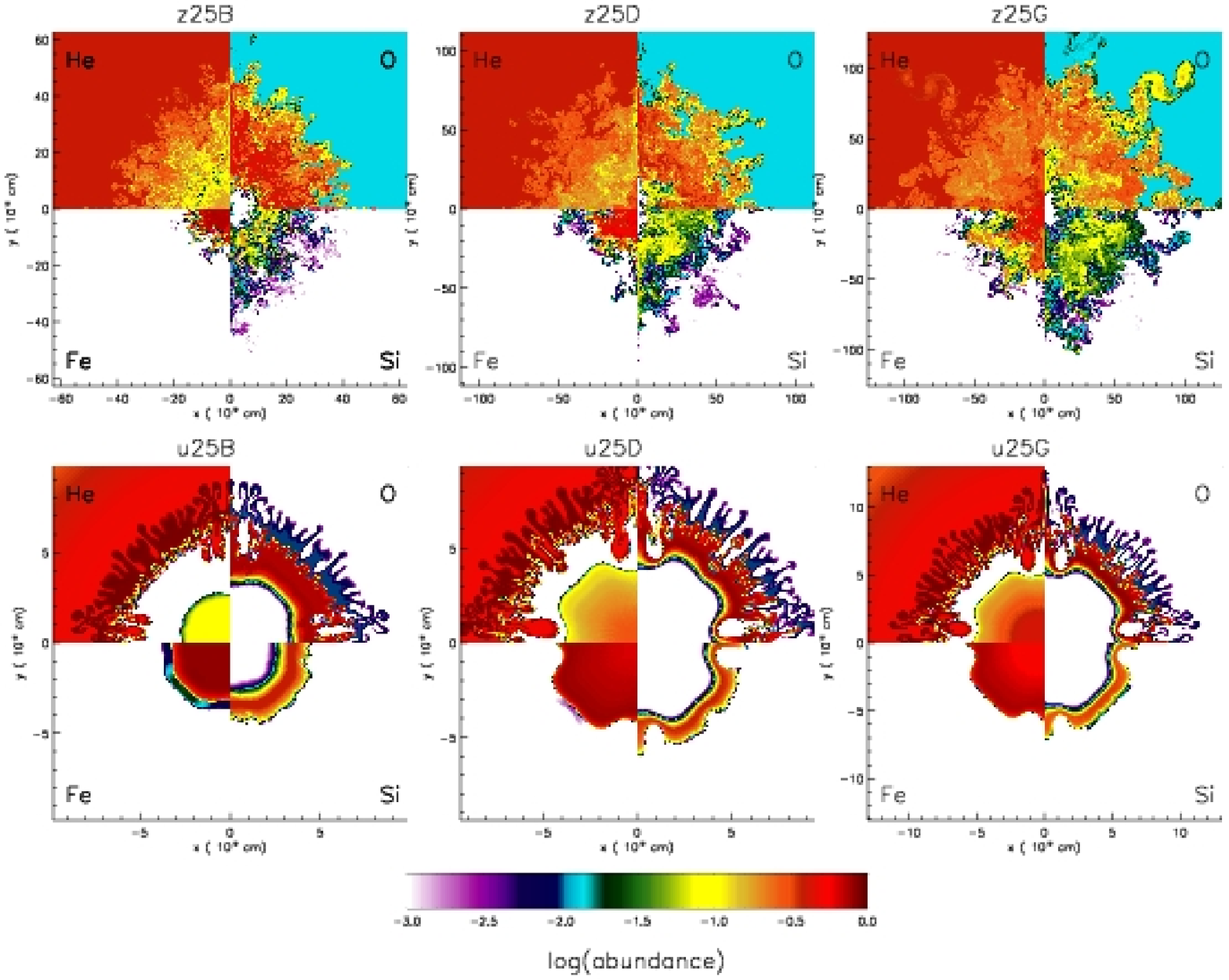}
  \includegraphics[height=.30\textheight]{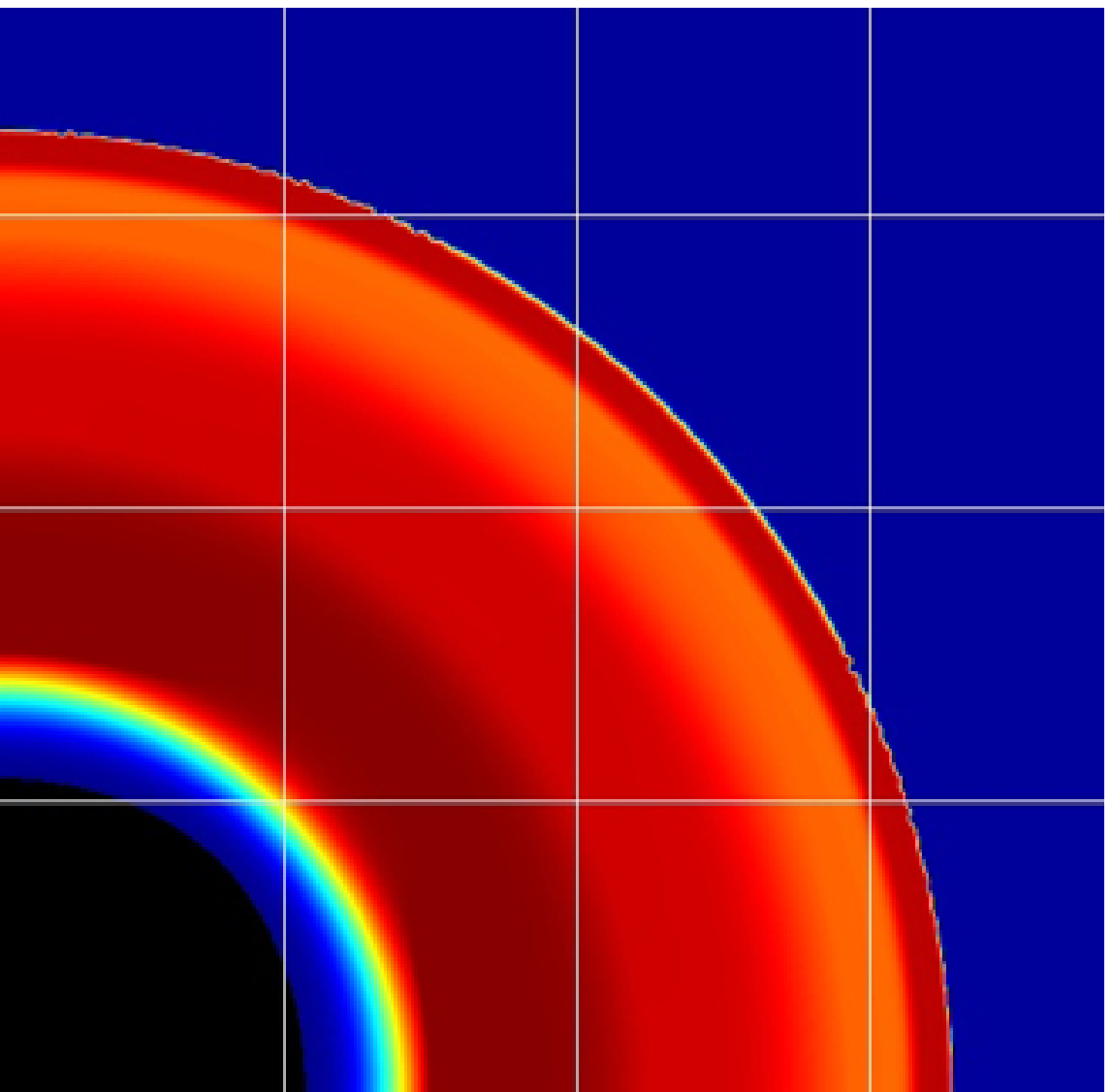}
  \caption{Mixing in primordial supernovae prior to shock breakout from 
   the surface of the star.  {\it Left}: core-collapse explosion of a 25 
   M$_{\odot}$ zero-metallicity progenitor. {\it Right}: a 200 M$_{\odot}$
   PISN.  Vigorous mixing of metals to high altitudes occurs in the CC SN
   but elements remain clearly segregated in the PISN.}
\end{figure}

%%%%%%%%%%%%%%%%%%%%%%%%%%%%%%%%%%%%%%%%%%%%%%%%
%% BACKMATTER
%%%%%%%%%%%%%%%%%%%%%%%%%%%%%%%%%%%%%%%%%%%%%%%%

\begin{theacknowledgments}

This work was carried out in part under the auspices of the National Nuclear 
Security Administration of the U.S. Department of Energy at Los Alamos National 
Laboratory supported by Contract No. DE-AC52-06NA25396. DW acknowledges support 
by the McWilliams Fellowship at the Bruce and Astrid McWilliams Center for 
Cosmology at Carnegie Mellon University.

\end{theacknowledgments}



\bibliographystyle{aipproc}   % if natbib is available

%%\bibliography{sample}

\IfFileExists{\jobname.bbl}{}
 {\typeout{}
  \typeout{******************************************}
  \typeout{** Please run "bibtex \jobname" to optain}
  \typeout{** the bibliography and then re-run LaTeX}
  \typeout{** twice to fix the references!}
  \typeout{******************************************}
  \typeout{}
 }

%%%%%%%%%%%%%%%%%%%%%%%%%%%%%%%%%%%%%%%%%%%
%% The following lines show an example how to produce a bibliography
%% without the help of the BibTeX program. This could be used instead
%% of the above.
%%%%%%%%%%%%%%%%%%%%%%%%%%%%%%%%%%%%%%%%%%%

\end{document}